# A Technique Based on Chaos for Brain Computer Interfacing

A. Banitalebi, S. K. Setarehdan and G. A. Hossein-Zadeh

Control and Intelligent Processing Centre of Excellence, Faculty of ECE, College of Engineering, University of Tehran

banitalebi@ieee.org, ksetareh@ut.ac.ir and ghzadeh@ut.ac.ir

*Abstract*—A user of Brain Computer Interface (BCI) system must be able to control external computer devices with brain activity. Although the proof-of-concept was given decades ago, the reliable translation of user intent into device control commands is still a major challenge. There are problems associated with classification of different BCI tasks. In this paper we propose the use of chaotic indices of the BCI. We use largest Lyapunov exponent, mutual information, correlation dimension and minimum embedding dimension as the features for the classification of EEG signals which have been released by BCI Competition IV. A multi-layer Perceptron classifier and a KM-SVM(support vector machine classifier based on k-means clustering) is used for classification process, which lead us to an accuracy of 95.5%, for discrimination between two motor imagery tasks.

## I. INTRODUCTION

The automatic and online translation of the user intent to the computer language is the mission of brain computer interfacing. This intent is estimated from the Electroencephalogram (EEG) measurements.

Several linear and nonlinear techniques have been applied to the EEG signals for BCI applications in the past[2, 3, ..., 12]. For example [2, 3] review the classification algorithms used in BCI, and some linear techniques are employed in [5, 6]. [7] used the wavelet transform and [8] proposed the MVAR model. [9] used independent component analysis (ICA) for BCI-EEG recognition. Also some nonlinear and chaotic methods are presented in [10, 11, 12].

Linear methods of EEG analysis such as Fourier transform or power spectral density, in comparison to chaotic analysis, are more computationally efficient but less strong in the interpretation of results [13, 14]. For example, [15] shows that through nonlinear time series analysis, one can discriminate even high-dimensional chaos from colored noise. Some of traditional linear methods have been found largely insensitive to task conditions associated with different brain dynamics [16, 17].

Recent researches on the human EEG, revealed the chaotic nature of this signal. In this research we take the advantage of this chaotic behavior in order to classify the EEG signals of a BCI task.

In section II the important indices are introduced briefly. Section III contains a short explanation of the EEG signals which we have used for simulation. In section IV is dedicated to the proposed method. Results are presented in section V, and section VI is conclusion.

## II. CHAOTIC INDICES

The chaotic indices used in this research are introduced in the following subsections.

### A. Largest Lyapunov Exponent

The Lyapunov exponent is a quantitative measure of the dynamics of trajectory evolution in the phase space. It capsulizes the average rate of convergence or divergence of two neighboring trajectories in the phase space. It can be negative, zero or positive. Negative values mean that the two trajectories draw closer to one another. Positive exponents on the other hand, result from the trajectory divergence and appear only within the chaotic domain. In other words, a positive Lyapunov exponent is one of the chaos indicators. Here we use the popular Wolf et al. [18] method for the calculation of largest Lyapunov exponent. It is summarized in the following equation:

$$l_1 = \lim_{n \to \infty} \frac{1}{n} \sum_{i=0}^{n-1} \log_e |f'(x_i)| \qquad (1)$$

In this equation, $f(x_i)$ is the local slop of trajectory. Because it averages local divergences and/or convergences from many places over the entire attractor, a *Lyapunov exponent is a global quantity*, not a local quantity.

Because the neighboring trajectories represent changes in initial conditions of a system, $l_1$ is an average or global measure of the sensitivity of the system to slight changes or perturbations. A system isn't sensitive at all in the non-chaotic regime, since any two nearby trajectories converge. In contrast, a system is highly sensitive in the chaotic regime, in that two neighboring trajectories separates, sometimes rapidly.

B. *Mutual Information*

If we denote $X, Y$ as two random variables, then $H_X, H_Y$ are their entropies and we have:

$$H_Y = -\sum_{j=1}^{N_s} p(y_j) \log_2 p(y_j) \qquad (2)$$

in which $N_s$ is the number of non-zero probabilities. The mutual information for $X, Y$ is defined as:

$$I_{Y;X} = H_Y + H_X - H_{X;Y} \qquad (3)$$

in which $H_{X;Y}$ is defined as:

$$H_{X;Y} = -\sum_{i=1}^{N_S} \sum_{j=1}^{N_S} p(x_i, y_j) \log_2 p(x_i, y_j) \qquad (4)$$

After substituting (4) in (3) and by some mathematical simplifications, we will have:

$$I_{Y;X} = \sum_{i=1}^{N_S} \sum_{j=1}^{N_S} p(x_i, y_j) \log_2 \frac{p(x_i, y_j)}{p(x_i) p(y_j)} \qquad (5)$$

For the calculation of $I_{Y;X}$ in EEG signals, in lag space $x_i$ becomes $x_t$ and $x_j$ becomes $x_{t+m}$. Bigger quantity of mutual information results in a less chaotic system. More details are reported in [19].

C. *Minimum Embedding Dimension*

For computational costs, simplicity of interpretation and other reasons, we'd like to reconstruct an attractor in a *small* embedding dimension. There's no theory or even a rule-of-thumb available in this regard. To solve the problem of false neighbor method, Cao proposed a method to choose the threshold value, which is often used to determine the embedding dimension. Different time series data may have different threshold values. These imply that it is difficult to give an independent reasonable threshold value which is independent of the dimension $d$ and each trajectory point, as well as the considered time series data. In this method a new quantity is defined:

$$E(m) = \frac{1}{N - mt} \sum_{i=1}^{N-mt} f(i, m) \qquad (6)$$

$E(m)$ is dependent only on the dimension $m$ and the lag $t$ and $f$ is a function of $i, m$. To investigate its variation from *m to m+1*, *E1(m)* is defined as:

$$E1(m) = \frac{E(m+1)}{E(m)} \qquad (7)$$

Cao found that *E1(m)* stops changing when $m$ is greater than some value $m_0$ if the time series comes from an attractor. Then $m_0 + 1$ is the minimum embedding dimension we look for. It is necessary to define another quantity which is useful to distinguish between deterministic and stochastic signals. Let

$$E2(m) = \frac{E^*(m+1)}{E^*(m)} \qquad (8)$$

where:

$$E^*(m) = \frac{1}{N - mt} \sum_{i=1}^{N-mt} |x_{i+mt} - x_{n(i,m)+mt}| \qquad (9)$$

More details are in [20].

D. *Correlation Dimension*

The correlation dimension is the most popular non-integer dimension currently used. It probes the attractor to a much finer scale than does the information dimension and is also easier and faster to compute. Like the information dimension, it takes into account the frequency with which the system visits different phase space zones. Most other dimensions involve moving a measuring device by equal, incremental lengths over the attractor. In contrast, the correlation dimension involves systematically locating the device at each datum point, in turn. The procedure usually begins by embedding the data in a two-dimensional pseudo phase space. For a given radius $e$, count the number of points within distance $e$ from the reference point. After doing that for each point on the trajectory, sum the counts and normalize the sum. That yields a correlation sum. Then repeat that procedure to get correlation sums for larger and larger values of $e$. A log plot of correlation sum versus $e$ (for that particular embedding dimension) typically shows a straight or nearly straight central region. The slop of that straight segment is the correlation dimension. For chaotic data, the correlation dimension initially increases with embedding dimension, but eventually (at least in the ideal case) it asymptotically approaches a final (true) value. We use Grassberger and Procaccia's [21] method for computing the correlation dimension $D_2$. More details are in [23].

III. THE EEG DATA

EEG signals were obtained from BCI Competition 2008 databank [24]. Scalp EEG data were recorded (with sampling rate of 1000 Hz) from healthy subjects during a motor imagery task without feedback. In this task. For each subject two classes of motor imagery were selected from the three classes *left hand*, *right hand*, and *foot* (side chosen by the subject; optionally also both feet).

A. *Calibration Data*

In the first two runs, arrows pointing left, right, or down were presented as visual cues to the subject on a computer screen. Cues were displayed for a period of 4 seconds during which the subject was instructed to perform the cued motor imagery task. These periods were interleaved with 2 seconds of blank screen and 2 seconds with a fixation cross shown in the centre of the screen. The fixation cross was superimposed on the cues, i.e. it was shown for 6 seconds.

*B. Evaluation Data*

Then 4 runs followed which are used for evaluating the submissions to the BCI competitions. Here, the motor imagery tasks were cued by soft acoustic stimuli (words *left*, *right*, and *foot*) for periods of varying length between 1.5 and 8 seconds. The end of the motor imagery period was indicated by the word *stop*. Intermitting periods had also a varying duration of 1.5 to 8 seconds. Note that in the evaluation data, there are not necessarily equally many trials from each condition.

Each of Healthy EEG's is recorded with a sampling frequency of 1000 Hz, from 59 EEG channels. More information about the dataset can be found in [30].

## IV. METHOD

A window with an adaptive width is used for the analysis of each EEG signal. The length of window is equal to the duration of task conditions. For all the signals (and so all the task trials), the chaos indices including largest Lyapunov exponent, mutual information, correlation dimension and minimum embedding dimension were calculated (for all windows of the signals). To simplify the calculations two matrixes were synthesized by concatenation of indices corresponding to different trials and subjects for test and training data. The dimension of the first matrix is 4*400*2 for the training set and that of second one is 4*450*2 for the test set. Each row corresponds to a feature (chaotic index that is computed for one trial), each column shows a single BCI task and the third dimension denotes the class of the task (which can be 1 or -1).

## V. RESULTS

Table 1, summarizes the mean and standard deviation (STD) of various indices in different tasks (for the whole test and train data). Although some of indices seem to be different between different tasks, but one must consider the variability of them inside each task.

As a complete illustration, Fig. 1 shows the relative frequency of the largest Lyapunov exponent in two tasks (Both are normalized). The curve with dots corresponds to task 1. Thus these curves may be interpreted as probability density functions (*pdf*) of largest Lyapunov exponent for the two tasks. Fig. 2 compares mutual information between the two states, Fig. 3 does the same for correlation dimension and Fig. 4 compares minimum embedding dimension between them. The rest of this section is dedicated to explaining the classification procedure. Fig. 5 shows the whole procedure of extracting the indices and classifying them.

TABLE I.
CHAOS INDICES IN DIFFERENT BCI TASKS, CALCULATED FROM EEG OF 10 SUBJECTS

|  | Class 1 = Task 1 | Class -1 = Task -1 |
|---|---|---|
|  | Mean (STD) | mean (STD) |
| largest Lyapunov exponent | 0.3821 (0.1246) | 0.6811 (0.1298) |
| mutual information | 0.4672 (0.0810) | 0.7741 (0.0907) |
| minimum embedding dimension | 0.5601 (0.1139) | 0.8596 (0.1221) |
| correlation dimension | 0.2364 (0.1797) | 0.5258 (0.1676) |

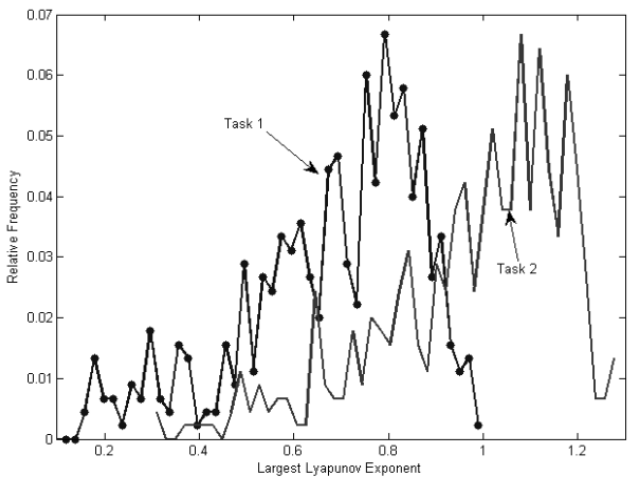

Figure 1. Comparison of largest Lyapunov exponent between tasks

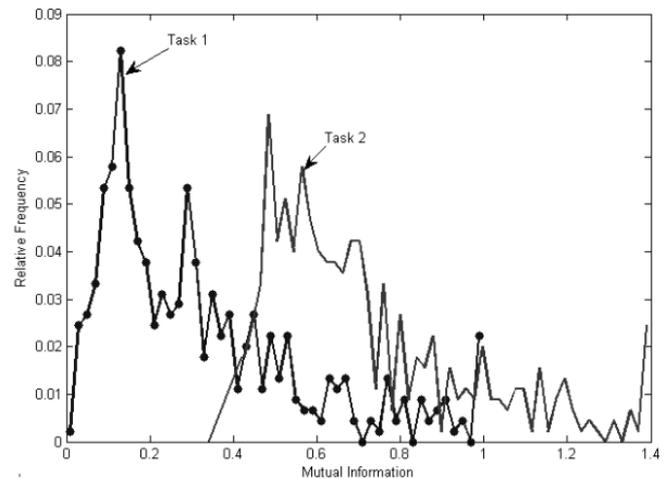

Figure 2. Comparison of mutual Information exponent between tasks

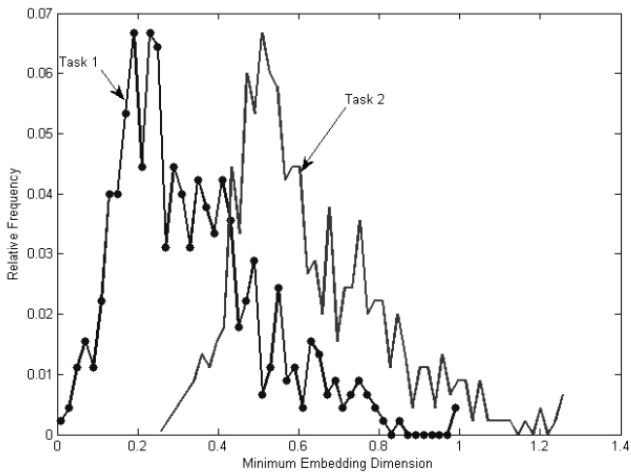

Figure 3. Comparison of minimum embedding dimension between tasks

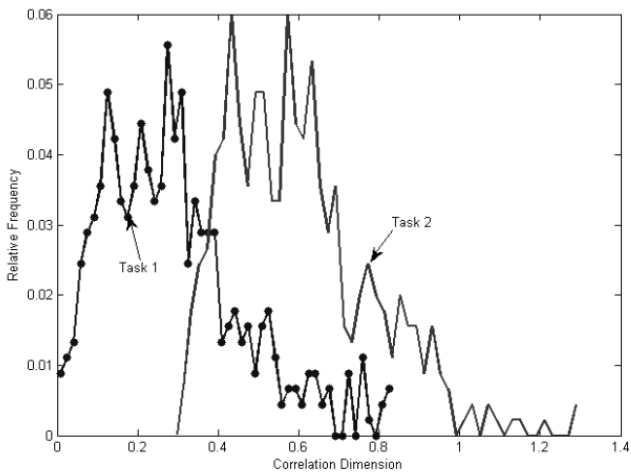

Figure 4. Comparison of correlation dimension between tasks

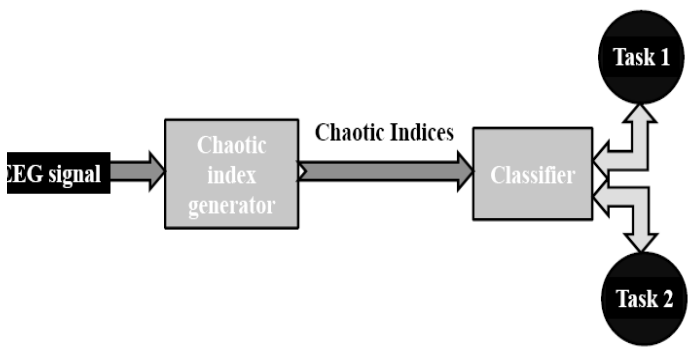

Figure 5. Diagram of the whole procedure

### A. Classification

We applied two classifying algorithms to the chaotic indices used here and demonstrated that the chaotic indices are excellent features for an accurate classification of brain computer interfacing EEG. The data (described in Section 3) is divided into a set of test data and a set of train data. 53% of the whole data samples assigned for test, which approximately equal to 900 windows. The train data were 47% of the whole data, and correspond to 800 windows. We also applied a feature conditioning process using PCA (principal component analysis) through SVD (singular value decomposition) for this classification.

### B. Multi-layer Perceptron Classifier

The first classifier we used was MLP classifier which belongs to the artificial neural networks family. MLP classifier includes four input neurons, one hidden layer including three neurons, and one output neuron. We utilized an adaptive conjugate gradient algorithm for the required optimization. The test and train procedure were both rapid and the simulation was quick. Correct classification rate was 95.1%. Also the classification percentage for the train data was nearly 98%.

### C. KM-SVM Classifier

The KM-SVM classifier (support vector machine classifier based on k-means clustering) is faster than MLP classifier and Bayesian classifiers, and it can be more accurate if the number of clusters had been chosen appropriately. First a k-means clustering algorithm is applied to the train dataset. Number of clusters in this procedure can be arbitrary, but we proposed to use 4 clusters. After that a SVM classifier is constructed using the cluster centers as the training data. The test data is then classified by this SVM classifier. More details of this classifier can be found in [31]. Construction of the classifier by use of only cluster centers leads to a rapid training procedure. The correct classifying percentage using this classifier is 95.5%. This supports the results obtained in this section, which was plotted in Figs 1, 2, 3 and 4. Regarding the good separation of histograms in these figures, we expect the corresponding features to be well suited for a classification study. The confusion matrix of the classifier is presented in table II. As we see from the table, diagonal elements of the table are near to 450, which is the whole number of trails in each class. For example there are 435 trials that really are in class 1 and the classifier classifies them truly. Other 15 trials are classified to be in class 2.

The *mse* value (Minimum Squared Error) using this classifier is 0.1788, which is a smaller value than all of the *mse* values obtained by the competitors in BCI competition 2008 (The results can be seen in the BCI competition IV website). As there were described in the competition website, a major number of the algorithms are based on the use of linear and non-dynamic features; This reality shows that chaotic features of the EEG signal in comparison to linear features can be more powerful in BCI classifications.

TABLE II.
CONFUSION MATRIX OBTAINED BY KM-SVM CLASSIFIER FOR DIFFERENT BCI TASKS

| Decided Class / Real Class | Class 1= Task 1 | Class 2= Task 2 |
|---|---|---|
| Class 1= Task 1 | 435 | 15 |
| Class 2= Task 2 | 25 | 425 |

## VI. CONCLUSION

In this paper we calculated several indices of chaos for BCI EEG signals of 10 healthy subjects. We used largest Lyapunov exponent, mutual information, correlation dimension and minimum embedding dimension as the features for the classification of EEG signals. Then we illustrated the usefulness of these indices through the classification. The MLP and KM-SVM were proposed as classifiers. Classifiers were established on four features, and successfully classified the tasks into two classes. In conclusion, it was shown that the four chaotic indices can separate the two classes with an accuracy of 95.5%.

REFERENCES

[1] A. Banitalebi, G. A. Hossein-Zadeh, "On the Variability of Chaos Indices in Sleep EEG Signals," *15th Iranian International Conference on Biomedical Engineering*, Mashhad, Iran, Feb. 2009.
[2] F. Lotte, M. Congedo, A. Lecuyer, F. Lamarche, B. Arnaldi, "A Review of Classification Algorithms for EEG-based Brain Computer Interfaces", *Journal of Neural Engineering*, March 2007.
[3] A. Akrami, S. Solhjoo, A. M. Nasrabadi, M. R. H. Golpayegani, "EEG-based mental task classification: linear and nonlinear classification of movement imagery", *Proceedings of the 2006 IEEE Engineering in Medicine and Biology, 27th Annual Conference*, Shanghai, China, Sep., 2006.
[4] J. R. Wolpaw, N. Birbaumer, W.J. Heetderks, D. J. McFarland, P. H. Peckham, G. Schalk, E. Donchin, L. A. Quatrano, C. J. Robinson, T. M. Vaughan, "Brain-Computer interface technology, a review of the first international meeting", *IEEE transactions on Rehabilitation Engineering*, Vol. 8, No. 2, pp. 164-173, June 2000.
[5] R. Boostani, B. Graimann, M. H. Moradi, G. Pfurtscheller, "A Comparison Approach Toward Finding the Best Feature and Classifier in Cue-Based BCI", *Med. Bio. Eng. Comput,* Vol. 45, pp. 403-412, Feb. 2007.
[6] D. P. Bruke, P. Kelly, P. de Chazal, R. B. Reilly, C. Finucane, "A Parametric Feature Extraction and Classification Strategy for Brain Computer Interfacing", *IEEE Transactions on Neural Systems and Rehabilitation Engineering*, Vol. 13, NO. 1, pp. 12-17, March 2006.
[7] J. Z. Xue, H. Zhang, C. X. Zheng, X. G. Yan, "Wavelet packet transform for feature extraction during mental tasks", *Proceedings of Second International Conference on Machine Learning and Cybernetics,* Xi'an, Nov. 2003.
[8] X. M. Pei, C. X. Zheng, "Feature extraction and classification of brain motor imagery task based on MVAR model", *Proceedings of Third International Conference on Machine Learning and Cybernetics,* Shanghai, Aug. 2004.
[9] C. I. Hung., P. L. Lee, Y. T. Wu, H. Y. Chen, L. F. Chen, L. F. Chen, T. C. Yeh, J. C. Hsieh, "Recognition of Motor Imagery EEG Using Independent Component Analysis and Machine Classifiers", *WSCG'04,* Plzen, Czech Republic, Feb., 2004.
[10] R. Boostani, M. H. Moradi, "A New Approach in BCI Research Based on Fractal Dimension as Feature and Adaboost as Classifier", *Journal of Neural Engineering*, Vol. 1, pp. 212-217, Nov. 2004.
[11] J. Zhang, G. Li, W. J. Freeman, "Application of Novel Chaotic Neural Networks to Text Classification Based on PCA", *LNCS4319, Springer*, pp. 1041-1048, 2006.
[12] R. Hu, G. Li, M. Hu, J. Fu, W. J. Freeman, "Recognition of ECOG in BCI Systems Based on a Chaotic Neural Model", *LNCS4491, Springer*, pp. 685-693, 2007.
[13] G. Mayer-Kress, S.P. Layne, Dimensionality of the Human Electroencephalogram, in: S.H. Koslow (Ed.), *Perspectives in Biological Dynamics and Theoretical Medicine*, Annuls of New York Academy of Sciences, vol. 54, pp. 62-87, New York 1987.
[14] V. Muller, W. Lutzenberger, H. Preibl, F. Pulvermuller, N. Birbaumer, "Complexity of Visual Stimuli and Nonlinear EEG Dynamics in Humans", *Cognitive Brain Research*, vol. 16, pp. 104-110, 2003.
[15] U. Parlitz, "Nonlinear Time Series Analysis", *Proceeding of NDES'95*, Dublin, Ireland, pp. 28-29, 1995.
[16] W. Lutzenberger, "EEG Alpha Dynamics as Viewed from EEG Dimension Dynamics", *Int. J. Psychophysiol*, vol. 26, pp. 273–283, 1997.
[17] V. Muller, W. Lutzenberger, F. Pulvermuller, B. Mohr, N. Birbaumer, "Investigation of Brain Dynamics in Parkinson's Disease by Methods Derived from Nonlinear Dynamics", *Exp. Brain Res.* vol. 137, pp. 103–110, 2001.
[18] A. Wolf, J. B. Swift, H. L. Swinney, J. A. Vastano, "Determining Lyapunov Exponents from a Time Series". *Physica*, vol. 16, pp. 285-317, 1985.
[19] G. P. Williams, *Chaos Theory Tamed*, JOSEPH HENRY, chapter 24, 25, 27, 1997.
[20] L. Y. Cao, "Practical Method for Determining the Minimum Embedding Dimension of a Scalar Time Series", *Physica,* vol. 16, pp. 43-50, 1997.
[21] P. Grassberger, I. Procaccia, "Measuring the Strangeness of Strange Attractors", *Physica*, vol. 9, pp. 183-208, 1983.
[22] F. Takens, "Detecting Strange Attractors in Turbulance", *Dynamical Systems and Turbulence*, *Springer*, pp. 366-381, Berlin, 1981.
[23] R. Q. Quiroga, *"Quantitive Analysis of EEG Signals: Time-Frequency Methods and Chaos Theory"*, Institute of Physiology and Institute of Signal Processing Medical University of Lubeck, pp. 68-77, 1998.
[24] http://ida.first.fraunhofer.de/projects/bci/competition_iv/desc_1.html
[25] Douglas C. Montgomery, George C. Runger, *Applied statistics and probability for engineers*, John Wiley & Sons, 2002.
[26] John A. Kassebaum, Brian H. Foresman, Thomas M. Talavage, Russell C. Eberhart, "Observations from Chaotic Analysis of Sleep EEGs", *Proceedings of 28th IEEE EMBS Annual International Conference*. New York city, USA, pp. 2126-2129, 2006.
[27] R. Acharya, O. Faust, N. Kannathal, T. Chua, S. Laxminarayan, "Nonlinear Analysis of EEG Signals at Various Sleep Stages", *Computer Methods and Programs in Biomedicine*. pp. 37-45, 2005.
[28] K. Susmakova, "Human Sleep and Sleep EEG", *Measurement Science Review*. vol. 4, pp. 54-75, 2004.
[29] M. B. Kennel, R. Brown and H. D. I. Abarbanel, "Determining Embedding Dimension for Phase Space Reconstruction Using a Geometrical Construction", *Phys. Rev. A*, pp.34-45, 1992.
[30] B. Blankertz, G. Dornhege, M. Krauledat, K. R. Müller, and G. Curio, "The Non-invasive Berlin Brain-Computer Interface: Fast Acquisition of Effective Performance in Untrained Subjects", *NeuroImage*, 37(2):539-550, 2007.
[31] J. Wang, X. Wu and C. Zhang, "Support Vector Machines Based on K-means Clustering for Real-time Business Intelligence Systems," *Int. J. Business Intelligence and Data Mining*, Vol. 1, No. 1, pp.54–64, 2005.